\documentstyle[pra,aps]{revtex}
\begin{document}
\draft
\title{A scheme for the teleportation of multiqubit quantum information via the
control of many agents in a network}
\author{Chui-Ping Yang and Siyuan Han}
\address{Department of Physics and Astronomy, University of Kansas, Lawrence,\\
Kansas 66045, USA}
\maketitle

\begin{abstract}
We present a new scheme for teleporting multiqubit quantum information from
a sender to a distant receiver via the control of many agents in a network.
We show that the receiver can successfully restore the original state of
each qubit as long as all the agents cooperate. However, it is remarkable
that for certain type of teleported states, the receiver can not gain any
amplitude information even if one agent does not collaborate. In addition,
our analysis shows that for general input states of each message qubit, the
average fidelity for the output states, when even one agent does not take
action, is the same as that for the previous proposals.
\end{abstract}

\pacs{PACS number{s}: 03.67.Lx, 03.65.-w}
\date{\today }

%\narrowtext

In 1993 Bennett {\it et al}. showed that an unknown quantum state of a
two-state particle can be teleported from one place to another [1]. Since
then, much progress has been made in this area. Experimentally, quantum
teleportation was recently demonstrated using photon polarization states
[2], optical coherent states [3], and nuclear magnetic resonance [4]. On the
other hand, Karlsson and Bourennane [5] and Hillery {\it et. al.} [6]
generalized the idea of Bennett {\it et al.} to the ``one to many'' quantum
communication of teleportation via many-particle quantum channels, i.e.,
quantum secret sharing---splitting a message among many spatially-separated
agents in a network so that no subset of agents is sufficient to read the
message but the entire set is. Thereafter, a lot of works on quantum secret
sharing were presented [7-12].

Recently, there is an increasing interest in the controlled teleportation.
Different from the secret sharing, the controlled teleportation is actually
the ``one to one (a sender to a distant receiver)'' quantum information
transfer via the control of agents in a network. The controlled
teleportation is useful in the context of quantum information such as
networked quantum information processing and cryptographic conferencing
[13,14]. And also, it may have other interesting applications, such as in
opening a credit account on the agreement of all the managers in a network.
Recently, several schemes have been proposed [15,16] for controlled
teleportation. As relevant to this work, we may cite the works in Refs.
[5,6], and the recent work in Ref. [15] on simultaneous teleportation of
multiqubit quantum information via the control of many agents in a network.
The distinct feature of the scheme described in [15] is that compared with
the method in [5,6], the resources required for auxiliary qubits, local
operations, and classical communication are greatly reduced. However, as
addressed in [15], when the teleported message qubit $i$ is initially in an
arbitrary state $\alpha _i\left| 0\right\rangle +\beta _i\left|
1\right\rangle ,$ the resulting density operator for the corresponding qubit 
$i^{\prime \prime }$ belonging to the receiver would be 
\begin{equation}
\rho _{i^{\prime \prime }}=\left| \alpha _i\right| ^2\left| 0\right\rangle
\left\langle 0\right| +\left| \beta _i\right| ^2\left| 1\right\rangle
\left\langle 1\right| 
\end{equation}
or 
\begin{equation}
\rho _{i^{\prime \prime }}=\left| \alpha _i\right| ^2\left| 1\right\rangle
\left\langle 1\right| +\left| \beta _i\right| ^2\left| 0\right\rangle
\left\langle 0\right| 
\end{equation}
(depending on the sender's Bell-state measurement outcome), when one agent
does not collaborate. This result shows that the receiver can obtain
amplitude information about the sender's each message qubit even though
he/she knows nothing about its phase. We note that the methods in [5,6] also
have this shortcoming since the same results (1) and (2) can be obtained if
not all agents collaborate (for the details, see reference [15]). Therefore,
it is not secure to employ the previous methods to realize the controlled
teleportation provided information is encoded through amplitude.

In this letter we will present a scheme for realizing the simultaneous
teleportation of multiqubit quantum information from a sender to a distant
receiver via the control of many agents in a network. We will show that
using this scheme, the receiver can restore the original state of each
message qubit when all the agents cooperate. However, it is interesting to
note that for certain kinds of teleported states, the receiver cannot gain
any amplitude information even if one agent does not collaborate. In
addition, our analysis shows that for general input states of each message
qubit, the average fidelity of the output states, when even one agent does
not cooperate, is the same as that of the previous proposals [5,6,15].

Consider that Alice has a string of message qubits labeled by $1,2,...,m,$
which is initially in the state $\prod_{i=1}^m(\alpha _i\left|
0\right\rangle _i+\beta _i\left| 1\right\rangle _i).$ To simplify our
presentation, we denote $\left| \varphi _i\right\rangle \equiv \alpha
_i\left| 0\right\rangle _i+\beta _i\left| 1\right\rangle _i$ in the
following. Suppose that Alice wishes to send the $m$-qubit information to
Bob via the control of $n$ agents ($A_1,A_2,...,A_n$) in a network, such
that Bob can get the complete information of each message qubit only if all
the agents collaborate. This task can be implemented with the following
prescription:

Firstly, Alice prepares the following EPR-GHZ entangled state through local
logic gates 
\begin{equation}
\prod_{i=1}^m\left| \text{EPR}\right\rangle _{i^{\prime }i^{\prime \prime
}}\otimes \left| \text{GHZ}\right\rangle _{+}+\prod_{i=1}^m\widetilde{\left| 
\text{EPR}\right\rangle }_{i^{\prime }i^{\prime \prime }}\otimes \left| 
\text{GHZ}\right\rangle _{-},
\end{equation}
where $\left| \text{EPR}\right\rangle _{i^{\prime }i^{\prime \prime
}}=\left| 00\right\rangle _{i^{\prime }i^{\prime \prime }}+\left|
11\right\rangle _{i^{\prime }i^{\prime \prime }}$ and $\widetilde{\left| 
\text{EPR}\right\rangle }_{i^{\prime }i^{\prime \prime }}=\left|
01\right\rangle _{i^{\prime }i^{\prime \prime }}-\left| 10\right\rangle
_{i^{\prime }i^{\prime \prime }}$ are EPR states for the qubit pair ($%
i^{\prime },i^{\prime \prime }$), and $\left| \text{GHZ}\right\rangle _{\pm
}=\left| 0\right\rangle ^{\otimes n}\left| 0\right\rangle _a\pm \left|
1\right\rangle ^{\otimes n}$ $\left| 1\right\rangle _a$ are ($n+1$)-qubit
GHZ states. Then, Alice sends the first $n$ GHZ qubits to the $n$ agents and
the $m$ EPR qubits ($1^{\prime \prime },2^{\prime \prime },...,m^{\prime
\prime }$) to Bob, while keeping the last GHZ qubit (labeled by $a$) and the
other $m$ EPR qubits ($1^{\prime },2^{\prime },...,m^{\prime }$) to herself.
The state of the whole system is given by 
\begin{equation}
\ \ \ \ \ \prod_{i=1}^m\left| \varphi _i\right\rangle \left| \text{EPR}%
\right\rangle _{i^{\prime }i^{\prime \prime }}\otimes \left| \text{GHZ}%
\right\rangle _{+}+\prod_{i=1}^m\left| \varphi _i\right\rangle \widetilde{%
\left| \text{EPR}\right\rangle }_{i^{\prime }i^{\prime \prime }}\otimes
\left| \text{GHZ}\right\rangle _{-},
\end{equation}
which can be expressed as 
\begin{eqnarray}
&&\ \ \ \ \ \ \prod_{i=1}^m\left[ \left| \phi _{ii^{\prime
}}^{+}\right\rangle (\alpha _i\left| 0\right\rangle _{i^{\prime \prime
}}+\beta _i\left| 1\right\rangle _{i^{\prime \prime }})+\left| \phi
_{ii^{\prime }}^{-}\right\rangle (\alpha _i\left| 0\right\rangle _{i^{\prime
\prime }}-\beta _i\left| 1\right\rangle _{i^{\prime \prime }})\right. 
\nonumber \\
&&\ \ \ \ \ \ \left. +\left| \psi _{ii^{\prime }}^{+}\right\rangle (\alpha
_i\left| 1\right\rangle _{i^{\prime \prime }}+\beta _i\left| 0\right\rangle
_{i^{\prime \prime }})+\left| \psi _{ii^{\prime }}^{-}\right\rangle (\alpha
_i\left| 1\right\rangle _{i^{\prime \prime }}-\beta _i\left| 0\right\rangle
_{i^{\prime \prime }})\right]  \nonumber \\
&&\ \ \ \ \ \ \otimes \left| \text{GHZ}\right\rangle _{+}  \nonumber \\
&&\ \ \ \ \ \ +\prod_{i=1}^m\left[ \left| \phi _{ii^{\prime
}}^{+}\right\rangle (\alpha _i\left| 1\right\rangle _{i^{\prime \prime
}}-\beta _i\left| 0\right\rangle _{i^{\prime \prime }})+\left| \phi
_{ii^{\prime }}^{-}\right\rangle (\alpha _i\left| 1\right\rangle _{i^{\prime
\prime }}+\beta _i\left| 0\right\rangle _{i^{\prime \prime }})\right. 
\nonumber \\
&&\ \ \ \ \ \ \left. +\left| \psi _{ii^{\prime }}^{+}\right\rangle (-\alpha
_i\left| 0\right\rangle _{i^{\prime \prime }}+\beta _i\left| 1\right\rangle
_{i^{\prime \prime }})+\left| \psi _{ii^{\prime }}^{-}\right\rangle (-\alpha
_i\left| 0\right\rangle _{i^{\prime \prime }}-\beta _i\left| 1\right\rangle
_{i^{\prime \prime }})\right]  \nonumber \\
&&\ \ \ \ \ \ \otimes \left| \text{GHZ}\right\rangle _{-}.
\end{eqnarray}

Above and below, the subscripts $ii^{\prime }=11^{\prime },22^{\prime
},33^{\prime }...;$ $i^{\prime }i^{\prime \prime }=1^{\prime }1^{\prime
\prime },2^{\prime }2^{\prime \prime },3^{\prime }3^{\prime \prime }...;$
and $i^{\prime \prime }=1^{\prime \prime },2^{\prime \prime },3^{\prime
\prime }...;$ for $i=1,2,3...$. In addition, normalization factors
throughout this letter are omitted for simplicity. The states $\left| \phi
_{ii^{\prime }}^{+}\right\rangle ,\left| \phi _{ii^{\prime
}}^{-}\right\rangle ,\left| \psi _{ii^{\prime }}^{+}\right\rangle ,$ and $%
\left| \psi _{ii^{\prime }}^{-}\right\rangle $ related to Eq. (5) are the
four Bell states for the qubit pair ($i,i^{\prime }$), taking the form of 
\begin{eqnarray}
\left| \phi _{ii^{\prime }}^{\pm }\right\rangle &=&\left| 00\right\rangle
_{ii^{\prime }}\pm \left| 11\right\rangle _{ii^{\prime }},  \nonumber \\
\left| \psi _{ii^{\prime }}^{\pm }\right\rangle &=&\left| 01\right\rangle
_{ii^{\prime }}\pm \left| 10\right\rangle _{ii^{\prime }}.
\end{eqnarray}

Secondly, Alice performs a series of two-qubit Bell-state measurements
respectively on $m$ qubit pairs ($1,1^{\prime }$), ($2,2^{\prime }$),...($%
m,m^{\prime }$). After that, one has

\begin{equation}
\left| \psi \right\rangle \left| \text{GHZ}\right\rangle _{+}+\left| \psi
^{\prime }\right\rangle \left| \text{GHZ}\right\rangle _{-},
\end{equation}
where $\left| \psi \right\rangle $ and $\left| \psi ^{\prime }\right\rangle $
are the states for the $m$ qubits ($1^{\prime \prime },2^{\prime \prime
},...,m^{\prime \prime }$) belonging to Bob, which are expressed as follows 
\begin{equation}
\left| \psi \right\rangle =\prod_{i=1}^m\left| \psi \right\rangle
_{i^{\prime \prime }},\qquad \left| \psi ^{\prime }\right\rangle
=\prod_{i=1}^m\left| \psi ^{\prime }\right\rangle _{i^{\prime \prime }}.
\end{equation}
Here, $\left| \psi \right\rangle _{i^{\prime \prime }}$ and $\left| \psi
\right\rangle _{i^{\prime \prime }}^{\prime }$ are the states of Bob's qubit 
$i^{\prime \prime }.$ From Eq. (5), it can be seen that if the qubit pair ($%
i,i^{\prime }$) is measured in the Bell states $\left| \phi _{ii^{\prime
}}^{+}\right\rangle ,$ $\left| \phi _{ii^{\prime }}^{-}\right\rangle ,$ $%
\left| \psi _{ii^{\prime }}^{+}\right\rangle ,$ and $\left| \psi
_{ii^{\prime }}^{-}\right\rangle ,$ then the states $\left| \psi
\right\rangle _{i^{\prime \prime }}$ and $\left| \psi \right\rangle
_{i^{\prime \prime }}^{\prime }$ are, respectively, given by 
\begin{eqnarray}
\left| \psi \right\rangle _{i^{\prime \prime }} &=&\alpha _i\left|
0\right\rangle _{i^{\prime \prime }}+\beta _i\left| 1\right\rangle
_{i^{\prime \prime }},\;\left| \psi ^{\prime }\right\rangle _{i^{\prime
\prime }}=\alpha _i\left| 1\right\rangle _{i^{\prime \prime }}-\beta
_i\left| 0\right\rangle _{i^{\prime \prime }},\,\,\text{\quad for }\left|
\phi _{ii^{\prime }}^{+}\right\rangle ,  \nonumber \\
\left| \psi \right\rangle _{i^{\prime \prime }} &=&\alpha _i\left|
0\right\rangle _{i^{\prime \prime }}-\beta _i\left| 1\right\rangle
_{i^{\prime \prime }},\;\left| \psi ^{\prime }\right\rangle _{i^{\prime
\prime }}=\alpha _i\left| 1\right\rangle _{i^{\prime \prime }}+\beta
_i\left| 0\right\rangle _{i^{\prime \prime }},\text{ \quad for }\left| \phi
_{ii^{\prime }}^{-}\right\rangle ,  \nonumber \\
\left| \psi \right\rangle _{i^{\prime \prime }} &=&\alpha _i\left|
1\right\rangle _{i^{\prime \prime }}+\beta _i\left| 0\right\rangle
_{i^{\prime \prime }},\;\left| \psi ^{\prime }\right\rangle _{i^{\prime
\prime }}=-\alpha _i\left| 0\right\rangle _{i^{\prime \prime }}+\beta
_i\left| 1\right\rangle _{i^{\prime \prime }},\text{ \ for }\left| \psi
_{ii^{\prime }}^{+}\right\rangle ,  \nonumber \\
\left| \psi \right\rangle _{i^{\prime \prime }} &=&\alpha _i\left|
1\right\rangle _{i^{\prime \prime }}-\beta _i\left| 0\right\rangle
_{i^{\prime \prime }},\;\left| \psi ^{\prime }\right\rangle _{i^{\prime
\prime }}=-\alpha _i\left| 0\right\rangle _{i^{\prime \prime }}-\beta
_i\left| 1\right\rangle _{i^{\prime \prime }},\text{ \ for }\left| \psi
_{ii^{\prime }}^{-}\right\rangle .
\end{eqnarray}
Eq. (9) demonstrates that according to Alice's Bell-state measurement
outcome for the pair ($i,i^{\prime }$), Bob can restore the original state $%
\alpha _i\left| 0\right\rangle _i+\beta _i\left| 1\right\rangle _i$ of the
message qubit $i$ from the state $\left| \psi \right\rangle _{i^{\prime
\prime }}$ or $\left| \psi ^{\prime }\right\rangle _{i^{\prime \prime }}$ of
his qubit $i^{\prime \prime }$, via a single-qubit Pauli operation $\sigma
_x,\sigma _y,$ or $\sigma _z$ on the qubit $i^{\prime \prime }.$

Thirdly, each agent and Alice perform a Hadamard transformation on their
respective GHZ qubits. After that, the state (7) becomes [15] 
\begin{equation}
\ \ \ \ \left| \psi \right\rangle \left[ \sum_{\left\{ x_l\right\} }\left|
\left\{ x_l\right\} \right\rangle \left| 0\right\rangle _a+\sum_{\left\{
x_l\right\} }\left| \left\{ y_l\right\} \right\rangle \left| 1\right\rangle
_a\right] +\left| \psi ^{\prime }\right\rangle \left[ \sum_{\left\{
x_l\right\} }\left| \left\{ x_l\right\} \right\rangle \left| 1\right\rangle
_a+\sum_{\left\{ y_l\right\} }\left| \left\{ y_l\right\} \right\rangle
\left| 0\right\rangle _a\right] ,
\end{equation}
where $\{x_l\}$ ($\left\{ y_l\right\} $) denotes $x_1x_2...x_n$ ($%
y_1y_2...y_n$) and $x_l,y_l\in \{0,1\},l=1,2,...,n$ for the $n$ GHZ qubits
belonging to the $n$ agents. Furthermore, $\sum_{\{x_l\}}\left|
\{x_l\}\right\rangle $ ($\sum_{\{y_l\}}\left| \{y_l\}\right\rangle $) sums
over all possible basis states $\left| \{x_l\}\right\rangle $ ($\left|
\{y_l\}\right\rangle $) each containing an $even$ ($odd$) number of ``1''s.
For instance, when $n=3,$ $\sum_{\{x_l\}}\left| \{x_l\}\right\rangle =\left|
000\right\rangle +\left| 110\right\rangle +\left| 101\right\rangle +\left|
011\right\rangle $ while $\sum_{\{y_l\}}\left| \{y_l\}\right\rangle =\left|
001\right\rangle +\left| 010\right\rangle +\left| 100\right\rangle +\left|
111\right\rangle .$

Lastly, each agent and Alice make a measurement on their respective GHZ
qubits, and then send their measurement results to Bob via classical
channels. Based on Eq. (10), it is easy to see that Bob can predict whether
his $m$ qubits ($1^{\prime \prime },2^{\prime \prime },...,m^{\prime \prime
} $) are in $\left| \psi \right\rangle $ or $\left| \psi ^{\prime
}\right\rangle $, according to whether the outcomes of the $n$ agents'
measurement on their $n$ GHZ qubits contain an $even$ or an $odd$ number of
``1''s as well as whether Alice measures her GHZ qubit in the state $\left|
0\right\rangle $ or $\left| 1\right\rangle $.

Note that the state $\left| \psi \right\rangle $ ($\left| \psi ^{\prime
}\right\rangle $) is a product of individual-qubit states $\left| \psi
\right\rangle _{1^{\prime \prime }},\left| \psi \right\rangle _{2^{\prime
\prime }},...,\left| \psi \right\rangle _{m^{\prime \prime }}$ ($\left| \psi
^{\prime }\right\rangle _{1^{\prime \prime }},\left| \psi ^{\prime
}\right\rangle _{2^{\prime \prime }},...,\left| \psi ^{\prime }\right\rangle
_{m^{\prime \prime }}$) for the qubits ($1^{\prime \prime },2^{\prime \prime
},...,m^{\prime \prime }$). Hence, Bob can recover the original state $%
\alpha _i\left| 0\right\rangle _i+\beta _i\left| 1\right\rangle _i$ of
message qubit $i$ from the state $\left| \psi \right\rangle _{i^{\prime
\prime }}$ or $\left| \psi ^{\prime }\right\rangle _{i^{\prime \prime }}$ of
his qubit $i^{\prime \prime },$ according to the outcome of Alice's
Bell-state measurement on the qubit pair ($i,i^{\prime }$) and via a
single-qubit Pauli operation $\sigma _x,\sigma _y,$ or $\sigma _z$ on the
qubit $i^{\prime \prime }$ as described above. Therefore, the quantum
information originally carried by the $m$ message qubits ($1,2,...,m$) can
completely be restored by Bob, provided each agent performs a Hadamard
transformation followed by a measurement on his/her qubit.

Now let us turn to the next question: What will happen if $k$ agents do not
collaborate ($1\leq k<n)$ ? To answer it, let us return to the state (7).
Clearly, this state can be rewritten as 
\begin{eqnarray}
&&\ \ \ \ \left| \psi \right\rangle \left[ \left( \left| \phi
^{+}\right\rangle +\left| \phi ^{-}\right\rangle \right) \left|
0\right\rangle ^{\otimes k}+\left( \left| \phi ^{+}\right\rangle -\left|
\phi ^{-}\right\rangle \right) \left| 1\right\rangle ^{\otimes k}\right] 
\nonumber \\
&&\ \ \ \ +\left| \psi ^{\prime }\right\rangle \left[ \left( \left| \phi
^{+}\right\rangle +\left| \phi ^{-}\right\rangle \right) \left|
0\right\rangle ^{\otimes k}-\left( \left| \phi ^{+}\right\rangle -\left|
\phi ^{-}\right\rangle \right) \left| 1\right\rangle ^{\otimes k}\right] ,
\end{eqnarray}
where $\left| 0\right\rangle ^{\otimes k}$ and $\left| 1\right\rangle
^{\otimes k}$ are the two states of the GHZ qubits belonging to the $k$
agents who do not cooperate with Bob, while $\left| \phi ^{\pm
}\right\rangle =\left| 0\right\rangle ^{\otimes (n-k)}\left| 0\right\rangle
_a\pm \left| 1\right\rangle ^{\otimes (n-k)}\left| 1\right\rangle _a$ are
the GHZ states of the remaining $n-k+1$ GHZ qubits belonging to Alice and
the other $n-k$ agents who collaborate with Bob.

It is straightforward to show that if the other $n-k$ agents and Alice
perform a Hadamard transformation on their respective GHZ qubits, then the
states $\left| \phi ^{+}\right\rangle $ and $\left| \phi ^{-}\right\rangle $
are transformed into 
\begin{eqnarray}
\left| \phi ^{+}\right\rangle &\rightarrow &\sum_{\left\{ x_l^{\prime
}\right\} }\left| \left\{ x_l^{\prime }\right\} \right\rangle \left|
0\right\rangle _a+\sum_{\left\{ y_l^{\prime }\right\} }\left| \left\{
y_l^{\prime }\right\} \right\rangle \left| 1\right\rangle _a,  \nonumber \\
\left| \phi ^{-}\right\rangle &\rightarrow &\sum_{\left\{ x_l^{\prime
}\right\} }\left| \left\{ x_l^{\prime }\right\} \right\rangle \left|
1\right\rangle _a+\sum_{\left\{ y_l^{\prime }\right\} }\left| \left\{
y_l^{\prime }\right\} \right\rangle \left| 0\right\rangle _a,
\end{eqnarray}
where $\{x_l^{\prime }\}$ ($\left\{ y_l^{\prime }\right\} $) denotes $%
x_1^{\prime }x_2^{\prime }...x_{n-k}^{\prime }$ ($y_1^{\prime }y_2^{\prime
}...y_{n-k}^{\prime }$) with $x_l^{\prime },y_l^{\prime }\in
\{0,1\},l=1,2,...,n-k$ for the $n-k$ GHZ qubits belonging to the other $n-k$
agents. And, $\sum_{\{x_l^{\prime }\}}\left| \{x_l^{\prime }\}\right\rangle $
($\sum_{\{y_l^{\prime }\}}\left| \{y_l^{\prime }\}\right\rangle $) sums over
all possible basis states $\left| \{x_l^{\prime }\}\right\rangle $ ($\left|
\{y_l^{\prime }\}\right\rangle $) each containing an $even$ ($odd$) number
of ``1''s. Substituting $\left| \phi ^{\pm }\right\rangle $ by (12), the
state (11) becomes

\begin{eqnarray}
&&\ \ \left[ (\left| \psi \right\rangle +\left| \psi ^{\prime }\right\rangle
)\left| 0\right\rangle ^{\otimes k}+(\left| \psi \right\rangle -\left| \psi
^{\prime }\right\rangle )\left| 1\right\rangle ^{\otimes k}\right]
\sum_{\left\{ x_l^{\prime }\right\} }\left| \left\{ x_l^{\prime }\right\}
\right\rangle \left| 0\right\rangle _a  \nonumber \\
&&\ \ +\left[ (\left| \psi \right\rangle +\left| \psi ^{\prime
}\right\rangle )\left| 0\right\rangle ^{\otimes k}-(\left| \psi
\right\rangle -\left| \psi ^{\prime }\right\rangle )\left| 1\right\rangle
^{\otimes k}\right] \sum_{\left\{ x_l^{\prime }\right\} }\left| \left\{
x_l^{\prime }\right\} \right\rangle \left| 1\right\rangle _a  \nonumber \\
&&\ \ +\left[ (\left| \psi \right\rangle +\left| \psi ^{\prime
}\right\rangle )\left| 0\right\rangle ^{\otimes k}-(\left| \psi
\right\rangle -\left| \psi ^{\prime }\right\rangle )\left| 1\right\rangle
^{\otimes k}\right] \sum_{\left\{ y_l^{\prime }\right\} }\left| \left\{
y_l^{\prime }\right\} \right\rangle \left| 0\right\rangle _a  \nonumber \\
&&\ \ +\left[ (\left| \psi \right\rangle +\left| \psi ^{\prime
}\right\rangle )\left| 0\right\rangle ^{\otimes k}+(\left| \psi
\right\rangle -\left| \psi ^{\prime }\right\rangle )\left| 1\right\rangle
^{\otimes k}\right] \sum_{\left\{ y_l^{\prime }\right\} }\left| \left\{
y_l^{\prime }\right\} \right\rangle \left| 1\right\rangle _a,
\end{eqnarray}
which indicates that if the other $n-k$ agents and Alice perform a
measurement on their respective GHZ qubits, the $m$ qubits ($1^{\prime
\prime },2^{\prime \prime },...,m^{\prime \prime }$) belonging to Bob will
be entangled with the $k$ GHZ qubits belonging to the $k$ agents who did not
cooperate with Bob.

It is easy to see from Eq. (13) that for every outcome $\left| \left\{
x_l^{\prime }\right\} \right\rangle \left| 0\right\rangle _a,\left| \left\{
x_l^{\prime }\right\} \right\rangle \left| 1\right\rangle _a,\left| \left\{
y_l^{\prime }\right\} \right\rangle \left| 0\right\rangle _a,$ or $\left|
\left\{ y_l^{\prime }\right\} \right\rangle \left| 1\right\rangle _a$ of the
other $n-k$ agents' and Alice's measurements on their GHZ qubits, the
density operator of Bob's $m$ qubits ($1^{\prime \prime },2^{\prime \prime
},...,m^{\prime \prime }$) would be given by 
\begin{equation}
\rho =\left| \psi \right\rangle \left\langle \psi \right| +\left| \psi
^{\prime }\right\rangle \left\langle \psi ^{\prime }\right| ),
\end{equation}
after tracing over the $k$ GHZ qubits belonging to the $k$ agents who did
not cooperate with Bob.

Based on Eqs. (8) and (9), it can be seen that in the case when the qubit
pair ($t,t^{\prime }$) is measured in the Bell states $\left| \phi
_{tt^{\prime }}^{+}\right\rangle ,$ $\left| \phi _{tt^{\prime
}}^{-}\right\rangle ,$ $\left| \psi _{tt^{\prime }}^{+}\right\rangle ,$ and $%
\left| \psi _{tt^{\prime }}^{-}\right\rangle ,$ the states $\left| \psi
\right\rangle $ and $\left| \psi ^{\prime }\right\rangle $ involved in Eq.
(14) can be expressed as follows: 
\begin{eqnarray}
\left| \psi \right\rangle &=&\left| \widetilde{\psi }\right\rangle (\alpha
_t\left| 0\right\rangle _{t^{\prime \prime }}+\beta _t\left| 1\right\rangle
_{t^{\prime \prime }}),\;\left| \psi ^{\prime }\right\rangle =\left| 
\widetilde{\psi }^{\prime }\right\rangle (\alpha _t\left| 1\right\rangle
_{t^{\prime \prime }}-\beta _t\left| 0\right\rangle _{t^{\prime \prime }}),%
\text{ \ \thinspace \thinspace \ for }\left| \phi _{tt^{\prime
}}^{+}\right\rangle ,  \nonumber \\
\left| \psi \right\rangle &=&\left| \widetilde{\psi }\right\rangle (\alpha
_t\left| 0\right\rangle _{t^{\prime \prime }}-\beta _t\left| 1\right\rangle
_{t^{\prime \prime }}),\;\left| \psi ^{\prime }\right\rangle =\left| 
\widetilde{\psi }^{\prime }\right\rangle (\alpha _t\left| 1\right\rangle
_{t^{\prime \prime }}+\beta _t\left| 0\right\rangle _{t^{\prime \prime }}),%
\text{ \ \ \thinspace \thinspace for }\left| \phi _{tt^{\prime
}}^{-}\right\rangle ,  \nonumber \\
\left| \psi \right\rangle &=&\left| \widetilde{\psi }\right\rangle (\alpha
_t\left| 1\right\rangle _{t^{\prime \prime }}+\beta _t\left| 0\right\rangle
_{t^{\prime \prime }}),\;\left| \psi ^{\prime }\right\rangle =\left| 
\widetilde{\psi }^{\prime }\right\rangle (-\alpha _t\left| 0\right\rangle
_{t^{\prime \prime }}+\beta _t\left| 1\right\rangle _{t^{\prime \prime }}),%
\text{ \ for }\left| \psi _{tt^{\prime }}^{+}\right\rangle ,  \nonumber \\
\left| \psi \right\rangle &=&\left| \widetilde{\psi }\right\rangle (\alpha
_t\left| 1\right\rangle _{t^{\prime \prime }}-\beta _t\left| 0\right\rangle
_{t^{\prime \prime }}),\;\left| \psi ^{\prime }\right\rangle =\left| 
\widetilde{\psi }^{\prime }\right\rangle (-\alpha _t\left| 0\right\rangle
_{t^{\prime \prime }}-\beta _t\left| 1\right\rangle _{t^{\prime \prime }}),%
\text{ \ for }\left| \psi _{tt^{\prime }}^{-}\right\rangle ,
\end{eqnarray}
where the subscript $t^{\prime \prime }$ indicates any one of the $m$ qubits
($1^{\prime \prime },2^{\prime \prime },...,m^{\prime \prime }$) belonging
to Bob, and the subscripts $t$ and $t^{\prime }$ denote Alice's message
qubit and her EPR qubit (corresponding to Bob's qubit $t^{\prime \prime }$),
respectively. Furthermore, the states $\left| \widetilde{\psi }\right\rangle 
$ and $\left| \widetilde{\psi }^{\prime }\right\rangle $ in Eq. (15) are the
states of the remaining $m-1$ qubits belonging to Bob (after excluding the
qubit $t^{\prime \prime }$), which are given by $\left| \widetilde{\psi }%
\right\rangle =\prod_k\left| \psi \right\rangle _{k^{\prime \prime }}$ and $%
\left| \widetilde{\psi }^{\prime }\right\rangle =\prod_k\left| \psi ^{\prime
}\right\rangle _{k^{\prime \prime }}$ ($k\neq t$). Here, $\left| \psi
\right\rangle _{k^{\prime \prime }}$ and $\left| \psi ^{\prime
}\right\rangle _{k^{\prime \prime }}$ are the states of qubit $k^{\prime
\prime }$ ($k^{\prime \prime }\neq t^{\prime \prime }$), which depend on the
outcome of Alice's Bell-state measurement on the pair ($k,k^{\prime }$) and
take the form as described by (9). From Eqs. (14) and (15), it can be shown
that for each outcome ($\left| \phi _{tt^{\prime }}^{+}\right\rangle ,\left|
\phi _{tt^{\prime }}^{-}\right\rangle ,\left| \psi _{tt^{\prime
}}^{+}\right\rangle ,$ or $\left| \psi _{tt^{\prime }}^{-}\right\rangle $)
of Alice's Bell-state measurement on the pair ($t,t^{\prime }$), after
tracing over the qubit $t^{\prime \prime },$ the density operator for Bob's
remaining $m-1$ qubits is given by 
\begin{equation}
\widetilde{\rho }=tr_{t^{\prime \prime }}(\rho )=\left| \widetilde{\psi }%
\right\rangle \left\langle \widetilde{\psi }\right| +\left| \widetilde{\psi }%
^{\prime }\right\rangle \left\langle \widetilde{\psi }^{\prime }\right| .
\end{equation}
Eq. (16) shows that the density operator, for Bob's remaining $m-1$
``non-traced'' qubits, has the same form as (14). Hence, repeating the above
single-qubit tracing procedure, we find that the density operator for any
qubit $i^{\prime \prime }$ belonging to Bob ($i^{\prime \prime }=1^{\prime
\prime },2^{\prime \prime },...,$ or $m^{\prime \prime }$) can, after
tracing over Bob's other $m-1$ qubits, be written as 
\begin{equation}
\rho _{i^{\prime \prime }}=\left| \psi \right\rangle _{i^{\prime \prime
}}\left\langle \psi \right| _{i^{\prime \prime }}+\left| \psi ^{\prime
}\right\rangle _{i^{\prime \prime }}\left\langle \psi ^{\prime }\right|
_{i^{\prime \prime }}.
\end{equation}
Based on Eqs. (9) and (17), it is obvious that when Alice measures her qubit
pair ($i,$ $i^{\prime }$) in either Bell state $\left| \phi _{ii^{\prime
}}^{+}\right\rangle $ or $\left| \psi _{ii^{\prime }}^{-}\right\rangle ,$
the density operator $\rho _{i^{\prime \prime }}$ is 
\begin{equation}
\rho _{i^{\prime \prime }}=\frac 12\left[ I+\left( \alpha _i\beta
_i^{*}-\alpha _i^{*}\beta _i\right) \left| 0\right\rangle \left\langle
1\right| -\left( \alpha _i\beta _i^{*}-\alpha _i^{*}\beta _i\right) \left|
1\right\rangle \left\langle 0\right| \right] ,
\end{equation}
where $I=\left| 0\right\rangle \left\langle 0\right| +\left| 1\right\rangle
\left\langle 1\right| $ is an identity. On the other hand, when Alice
measures the qubit pair ($i,$ $i^{\prime }$) in either Bell state $\left|
\phi _{ii^{\prime }}^{-}\right\rangle $ or $\left| \psi _{ii^{\prime
}}^{+}\right\rangle ,$ the density operator $\rho _{i^{\prime \prime }}$ is 
\begin{equation}
\rho _{i^{\prime \prime }}=\frac 12\left[ I-\left( \alpha _i\beta
_i^{*}-\alpha _i^{*}\beta _i\right) \left| 0\right\rangle \left\langle
1\right| +\left( \alpha _i\beta _i^{*}-\alpha _i^{*}\beta _i\right) \left|
1\right\rangle \left\langle 0\right| \right] .
\end{equation}
In Eqs. (18) and (19) the normalization factor $\frac 12$ is restored for
completeness. The above process demonstrates that the density operator (18)
or (19) depends only on the outcome of Alice's Bell-state measurement on the
pair ($i,$ $i^{\prime }$).

From Eqs. (18) and (19), one sees that the density operator $\rho
_{i^{\prime \prime }}$ would be $\frac 12I$ when $\alpha _i$ and $\beta _i$
are real. This means that when the message qubit $i$ is initially in the
state 
\begin{equation}
\cos \frac{\vartheta _i}2\left| 0\right\rangle _i+\sin \frac{\vartheta _i}2%
\left| 1\right\rangle _i,
\end{equation}
Bob cannot obtain any amplitude information from his $m$ qubits if there are 
$k$ agents who do not collaborate. One may worry about a relative phase
(dynamic phase) induced due to the free time evolution of the states $\left|
0\right\rangle $ and $\left| 1\right\rangle .$ However, this phase can be
avoided for certain kinds of physical qubits, such as polarized photons or
atoms/solid-state devices with degenerate ground levels $\left|
0\right\rangle $ and $\left| 1\right\rangle .$ It is well known that the
polarized photons are important qubit resources for linear optical quantum
computation and quantum communication [2,17-23]) and the use of qubits with
two degenerate ground levels $\left| 0\right\rangle $ and $\left|
1\right\rangle $ is an active way for fault-tolerant quantum computing.

If the initial state of the message qubit $i$ is of the general form 
\begin{equation}
\left| \varphi _i\right\rangle =\cos \frac{\vartheta _i}2\left|
0\right\rangle _i+e^{i\phi _i}\sin \frac{\vartheta _i}2\left| 1\right\rangle
_i,
\end{equation}
the fidelity of the state of qubit $i^{\prime \prime }$ would be 
\begin{eqnarray}
{\cal F}_{i^{\prime \prime }} &=&\left\langle \varphi _i\right| \rho
_{i^{\prime \prime }}\left| \varphi _i\right\rangle  \nonumber \\
&=&\frac 12\left( 1\pm \sin ^2\vartheta _i\sin ^2\phi _i\right)
\end{eqnarray}
when Alice measures the qubit pair ($i,$ $i^{\prime }$) in either of the
Bell states $\left| \phi _{ii^{\prime }}^{\pm }\right\rangle $ and $\left|
\psi _{ii^{\prime }}^{\mp }\right\rangle .$

To compare the present scheme with the methods in [5,6,15], it is necessary
to compute the fidelity averaged over $\vartheta _i$ and $\phi _i$ (assuming
random distributions). A simple calculation finds that the average fidelity
would be $\overline{{\cal F}}_{i^{\prime \prime }}=2/3$ if the qubit pair ($%
i,$ $i^{\prime }$) is measured in either $\left| \phi _{ii^{\prime
}}^{+}\right\rangle $ or $\left| \psi _{ii^{\prime }}^{-}\right\rangle .$ On
the other hand, the average fidelity would be $\overline{{\cal F}}%
_{i^{\prime \prime }}=1/3$ when the qubit pair ($i,$ $i^{\prime }$) is
measured in either $\left| \phi _{ii^{\prime }}^{-}\right\rangle $ or $%
\left| \psi _{ii^{\prime }}^{+}\right\rangle .$ However, this fidelity can
be raised to $2/3$ if Bob makes use of the result of Alice's Bell-state
measurement. This can be done in the following way. The density matrix of
Bob's qubit in Eq. (19) is for two of four results of Alice's Bell-state
measurement. If not all agents collaborate Bob does not know if he should
apply $\sigma _x$ or $\sigma _z$ operation. The best he can do is to
randomly choose between two possibilities. However, in both cases the result
is the same -- the density matrix of Eq. (19) is transformed into density
matrix of Eq. (18) for which the fidelity is $2/3.$ Therefore, the average
fidelity obtained in the present scheme is equal to that obtained using the
methods in [5,6,15]. One can verify based on Eqs. (1) and (2) that if the
methods in [5,6,15] are used, the average fidelity would be $\overline{{\cal %
F}}_{i^{\prime \prime }}=2/3,$ independent of the outcomes of the Bell-state
measurement for the qubit pair ($i,i^{\prime }$). We emphasize that the
result obtained above applies to the case of $k=1$. Namely, Bob can retrieve
all information only if all agents cooperate.

Several issues may need to be addressed here. Firstly, as a matter of fact,
the Hadamard transformations are not necessary for the present proposal. The
same results can be obtained when each agent performs a measurement on
his/her qubit in the basis $\left| +\right\rangle =\left| 0\right\rangle
+\left| 1\right\rangle $ and $\left| -\right\rangle =\left| 0\right\rangle
-\left| 1\right\rangle $ (instead of a Hadamard transformation and a
measurement in the basis $\left| 0\right\rangle $ and $\left| 1\right\rangle 
$)$,$ and then sends his/her measurement result $\left| +\right\rangle $ or $%
\left| -\right\rangle $ to the receiver. Secondly, the present proposal
actually does not require any specific order among Alice's Bell-state
measurement, her single-qubit operation, and each agent's operation. Lastly,
the present scheme is secure against eavesdropping and/or cheating as long
as Alice sends her classical information to Bob using standard quantum
cryptography [6,15]. For more discussions, we refer readers to reference
[15].

Before conclusion, it should be mentioned that for the present scheme, the
resources required for auxiliary qubits, local operations, and classical
communication{\it \ are the same as those needed by the protocol in [15] but
are dramatically reduced compared with the methods in [5,6]. }This is
because similar to the previous proposal in [15] the present scheme only
requires that: (i) the sender assigns one qubit to each agent; (ii) each
agent performs one single-qubit Hadamard transformation and one single-qubit
measurement on his/her qubit; and (iii) each agent sends one-bit classical
message to the receiver. {\it \ }For the detail discussion, see Ref. [15]%
{\it .}

In summary, we have presented a new scheme for teleporting multiqubit
quantum information from a sender to a distant receiver, via the control of
many agents in a network. We have shown that for general initial states of
each message qubit of arbitrary type, the average fidelity of the output
states of the receiver's qubit, when even one agent does not take any
action, is equivalent to that of the previous proposals [5,6,15]. Therefore,
as far as the control of each agent on teleportation for a general input
state, the present scheme is as efficient as the proposals [5,6,15].
However, as shown above, the present scheme has the distinct feature that
for certain kinds of teleported states (20) the receiver cannot gain any
amplitude information even if one agent does not collaborate. Note that the
states (20) are easily prepared especially for polarized photons (one can
modulate the amplitudes by a simple rotation of polarization). Therefore, we
believe that the present scheme is of great interest due to its more
security, as long as information is encoded through amplitudes of quantum
states.

This research was partially supported in part by National Science Foundation
ITR program (Grant No. DMR-0325551) and AFOSR (Grant No. F49620-01-1-0439),
funded under the Department of Defense University Research Initiative on
Nanotechnology (DURINT) Program and by the ARDA.

\end{document}